\documentclass[12pt]{spieman}  
\usepackage{amsmath,amsfonts,amssymb}
\usepackage{graphicx}
\usepackage{setspace}
\usepackage{tocloft}
\usepackage{makecell}
\usepackage{multirow}

\usepackage{lineno}
\usepackage{mathrsfs}

\title{Photon information efficiency limits in deep-space optical communications}

\author[a]{Marcin Jarzyna}
\author[a,b,*]{Ludwig Kunz}
\author[a]{Wojciech Zwoli\'{n}ski}
\author[a]{Micha{\l} Jachura}
\author[a,b,$\dagger$]{Konrad Banaszek}
\affil[a]{Centre for Quantum Optical Technologies, Centre of New Technologies, University of Warsaw, Banacha 2c, 02-097 Warsaw, Poland}
\affil[b]{Faculty of Physics, University of Warsaw, Pasteura 9, 02-093 Warsaw, Poland}

\cftpagenumbersoff{figure}
\cftpagenumbersoff{table} 
\begin{document} 
\maketitle


\begin{abstract}
Deep-space optical communication links operate under severely limited signal power, approaching the photon-starved regime which requires a receiver capable of measuring individual incoming photons. This makes the photon information efficiency (PIE), i.e. the number of bits that can be retrieved from a single received photon, a relevant figure of merit to characterize data rates achievable in deep-space scenarios. Here we review theoretical PIE limits assuming a scalable modulation format, such as pulse position modulation (PPM), combined with a photon counting direct detection receiver. For unrestricted signal bandwidth, the attainable PIE is effectively limited by the background noise acquired by the propagating optical signal. The actual PIE limit depends on the effectiveness of the noise rejection mechanism implemented at the receiver, which can be improved by the nonlinear optical technique of quantum pulse gating. Further enhancement is possible by resorting to photon number resolved detection, which improves discrimination of PPM pulses against weak background noise. The results are compared with the ultimate quantum mechanical PIE limit implied by the Gordon-Holevo capacity bound, which takes into account general modulation formats as well as any physically permitted measurement techniques. 
\end{abstract}


{\noindent \footnotesize\textbf{*}Present address: TNG Technology Consulting,
Beta-Stra\ss e 13a, 
85774 Unterf\"{o}hring, Germany}

{\noindent \footnotesize$^\dagger$Address all correspondence to Konrad Banaszek,  \linkable{k.banaszek@uw.edu.pl} }


\section{Introduction}
\begin{figure}[b]
\begin{center}
\begin{tabular}{c}
\includegraphics[width=12cm]{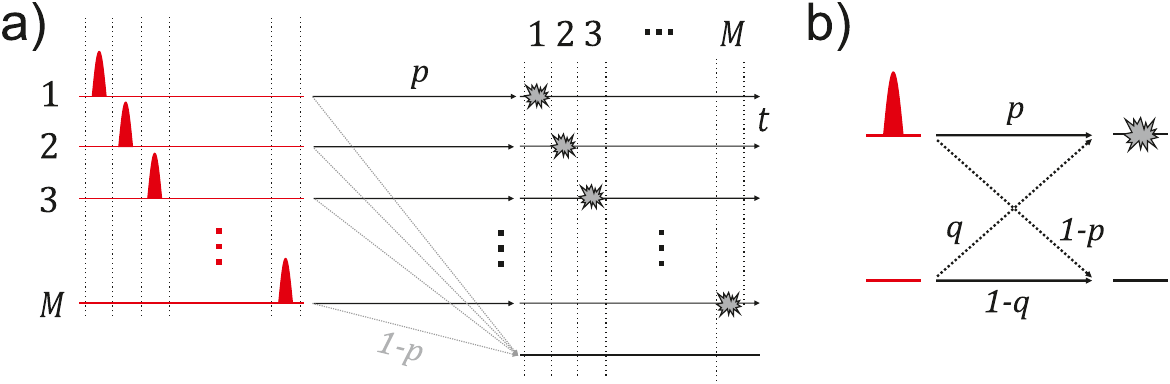}
\end{tabular}
\end{center}
\caption 
{ \label{Fig:PPM}
(a) Pulse position modulation (PPM) as a scalable modulation format. For each PPM symbol, a light pulse occupies a distinct temporal location in a frame of $M$ consecutive slots, thus encoding $\log_2 M$ bits of information. In an idealized scenario without background noise or detector dark counts, the timing of a count on a photon counting detector unambiguously identifies the received PPM symbol. Because of the uncertainty in the photon number for the incoming light pulse (given e.g.\ by the Poissonian statistics for laser light) there is a non-zero probability that no click will be generated over the entire frame, producing an erasure event. Increasing the signal bandwidth corresponds to shortening the duration of a single slot. (b) Conditional probabilities of generating a count on a detector for a pulse and an empty slot taking into account the presence of background noise.} 
\end{figure}

A widely recognized communication bottleneck in deep-space missions is the ability to transfer large volumes of data collected by onboard instruments back to Earth \cite{Hemmati2005,Hemmati2011,Biswas2017,Sodnik2017,LyrasIEEEAE2019,DeutschNatAstro2020,ArakiICSOS2022,RielanderICSO2022,GladyszICSO2023}.
Limitations of currently used radio frequency (RF) communication links can be lifted by redesigning communication systems to operate in the optical spectrum, which prospectively offers much higher data rates owing to wider bandwidths and lower diffraction losses in the course of signal propagation, as well as reduced regulatory requirements \cite{Williams2007}. Compared to the RF band, the much higher energy of a single quantum of the electromagnetic field at optical frequencies combined with power limitations implies an operating regime inherently distinct from that typical to conventional communication systems, either RF or fiber optic. 
Specifically, the operation of deep-space optical communication links easily reaches the photon-starved regime, when the average signal power spectral density is of the order of or even much less than the energy of a single photon at the carrier frequency per unit time-bandwidth area \cite{Boroson2018}. This implies the need to use photon-efficient modulation formats, such us pulse position modulation (PPM) shown in Fig.~\ref{Fig:PPM}(a), which encodes information in the location of a light pulse within a frame of otherwise empty temporal slots. The PPM format combined with direct detection (DD) implemented as time-resolved photon counting has become a standard in photon-starved scenarios \cite{Boroson2014,Biswas2018}.

There are several ways to improve the performance of the PPM/DD combination. The PPM order (i.e. the number of temporal slots per one PPM frame) and the frame duration may be optimized with respect to the received signal strength and the background noise strength, characterized by the respective average numbers of signal and noise photons per one slot.
The recently proposed and demonstrated in proof-of-principle settings technique of quantum pulse gating (QPG)\cite{Eckstein2011,BrechtReddyPRX2015,ReddyRaymerOpEx2017} can be employed to reduce the background noise by removing noise photons in temporal modes that do not match those of PPM pulses. As noted in Refs.~\citenum{Banaszek2019, Jarzyna2021, Jarzyna2023}, for low noise strengths (much below one photon per slot), further improvement is in principle possible by using photon number resolved (PNR) detection. 

The purpose of this paper is to discuss the performance of PPM/DD photon-starved links attainable in scenarios listed above using the photon information efficiency (PIE) as the figure of merit. This quantity conveniently allows one to convert the signal power reaching the receiver, expressed in terms of the received photon flux, i.e. the average number of photons per unit time integrated over the receive aperture, into the attainable data bit rate. It is also insightful to compare the results with the ultimate limits on the link performance derived from the quantum mechanical Gordon-Holevo (GH) capacity bound, which fully takes into account the dual wave-particle nature of the electromagnetic field and allows for the most general receivers compatible with the laws of quantum physics.\cite{Banaszek2020}



This paper is organized as follows. 
The relevance of the PIE to characterize the performance of power-limited free-space optical communication links is discussed in Sec.~\ref{Sec:PIE}. Quantum mechanical limits on the attainable PIE, including coherent detection scenarios and the ultimate Gordon-Holevo capacity bound are reviewed in Sec.~\ref{sec:pie_limits}. Next, Sec.~\ref{sec:PPM} presents theoretical results on optimizing the PPM order for a conventional DD receiver. Two possible enhancements of a DD receiver: noise reduction by means of QPG and improved discrimination of PPM pulses with PNR detection are studied in Sec.~\ref{Sec:Enhanced}. Obtained results are discussed in Sec.~\ref{Sec:Discussion}. Finally, Sec.~\ref{Sec:Conclusions} concludes the paper.

\section{Photon information efficiency (PIE)}
\label{Sec:PIE}

\begin{figure}
\begin{center}
\begin{tabular}{c}
\includegraphics[width=6.5cm]{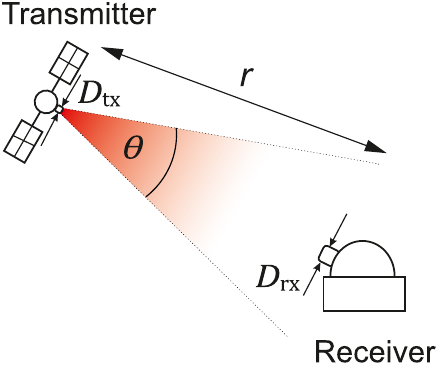}
\end{tabular}
\end{center}
\caption 
{ \label{fig:scheme}
A free-space optical communication link in a satellite-to-ground setting. $D_{\textrm{tx}}$, transmit aperture diameter; $D_{\textrm{rx}}$, receive aperture diameter; $r$, link range.} 
\end{figure} 

Consider a free-space optical communication link shown in Fig.~\ref{fig:scheme}. For convenience, relevant parameters are listed in Tab.~\ref{tab:parameters}. The relation between the received power $P_{\textrm{rx}}$ and the transmitted power $P_{\textrm{tx}}$ reads
\begin{equation}
P_{\textrm{rx}} = \eta_{\textrm{rx}} \cdot \eta_{\textrm{atm}} \cdot \eta_{\textrm{diff}} \cdot P_{\textrm{tx}},
\end{equation}
where $\eta_{\textrm{rx}}$ is the receiver subsystem efficiency, $\eta_{\textrm{atm}}$ is the atmospheric transmission, and $\eta_{\textrm{diff}}$ characterizes diffraction losses occurring in the course of propagation:
\begin{equation}
\eta_{\textrm{diff}} = \frac{1}{r^2} \cdot f_c^2 \cdot \left(
\frac{\pi D_{\textrm{rx}} D_{\textrm{tx}}}{4 c} \right)^2.
\end{equation}

Photon information efficiency $\textsf{PIE}$ is defined as the proportionality factor between the information rate $\textsf{R}$ and the received photon flux given by $P_{\textrm{rx}}/(hf_c)$. This yields the following expression in terms of the transmit power $P_{\textrm{tx}}$:
\begin{align}
    \textsf{R} & = \textsf{PIE} \cdot \frac{P_{\textrm{rx}}}{hf_c} \nonumber \\
    \label{Eq:RasPIE}
    & = \textsf{PIE} \cdot \frac{f_c}{h} \cdot \eta_{\textrm{rx}} \cdot \eta_{\textrm{atm}} \cdot \frac{1}{r^2} 
    \cdot \left(\frac{\pi D_{\textrm{rx}} D_{\textrm{tx}}}{4 c} \right)^2 \cdot P_{\textrm{tx}}.
\end{align}
Two observations are in place. First, the factor $1/r^2$ describes the inverse square scaling with the distance, as expected for power-limited free-space communication links. Second, the explicit scaling of the information rate $\textsf{R}$ with the carrier frequency $f_c$ is linear. This is a result of an interplay of two factors. On one hand, diffraction makes the power transmission factor scale quadratically with $f_c$, which is usually recalled as an argument in favor of the optical band versus the RF band. On the other hand however, for a given signal power the number of transmitted photons per unit time decreases inversely with the carrier frequency. Although in principle the PIE may depend implicitly on all the link parameters, 
we will see that without bandwidth (slot rate) limitations the PIE becomes predominantly a function of the background noise strength.

\begin{table}[t]
\caption{Optical link parameters.} 
\label{tab:parameters}
\begin{center}       
\begin{tabular}{|l|l|} 
\hline
\rule[-1ex]{0pt}{3.5ex}  Symbol & Description  \\
\hline\hline
\rule[-1ex]{0pt}{3.5ex}  $c$ & Speed of light  \\
\hline
\rule[-1ex]{0pt}{3.5ex}  $h$ & Planck's constant \\
\hline
\rule[-1ex]{0pt}{3.5ex}  $f_c$ & Carrier frequency  \\
\hline
\rule[-1ex]{0pt}{3.5ex}  $B$ & Slot rate (bandwidth)  \\
\hline
\rule[-1ex]{0pt}{3.5ex}  $r$ & Link range \\
\hline
\rule[-1ex]{0pt}{3.5ex}  $D_{\textrm{tx}}$ & Transmit aperture diameter   \\
\hline
\rule[-1ex]{0pt}{3.5ex}  $D_{\textrm{rx}}$ & Receive aperture diameter \\
\hline
\rule[-1ex]{0pt}{3.5ex}  $P_{\textrm{tx}}$ & Transmitted signal power  \\
\hline
\rule[-1ex]{0pt}{3.5ex}  $P_{\textrm{rx}}$ & Received signal power  \\
\hline
\rule[-1ex]{0pt}{3.5ex}  $\mathscr{N}$ & Background noise power   \\
\rule[-1ex]{0pt}{3.5ex}  & spectral density  \\
\hline
\end{tabular}
\end{center}
\end{table} 

\begin{figure}[h]
\begin{center}
\begin{tabular}{c}
\includegraphics[width=6.5cm]{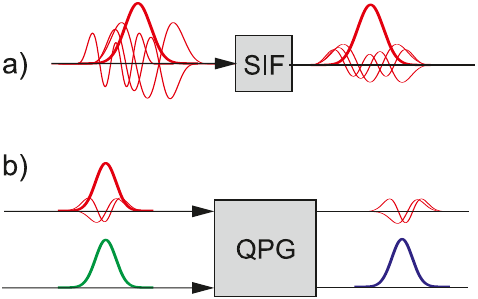}
\end{tabular}
\end{center}
\caption 
{ \label{Fig:Filtering}
(a) A sequential incoherent filter (SIF) applied to a temporally multimode input transmits the signal mode (thick line) while suppressing noise present in other temporal modes (thin lines). However, there is an inherent trade-off between the transmission coefficient of the signal mode and the suppression of noise in remaining modes. (b) Quantum pulse gating (QPG) relies on selective sum-frequency generation of the signal with a pump pulse (green line) to convert the signal mode to the sum-frequency band (blue line), while leaving noise present in other modes at the input frequency.} 
\end{figure}

In the following, the PIE will be analyzed as a function of two parameters that characterize respectively the strength of the received signal and the strength of the background noise. The signal strength is specified by the average number of received signal photons per slot:
\begin{equation}
n_s = \frac{1}{B} \cdot \frac{P_{\textrm{rx}}}{hf_c},
\label{Eq:nsdef}
\end{equation}
where the slot rate $B$ can be considered as a characterization of the signal bandwidth. The noise strength is quantified by the average number of background noise photons per slot $n_b$. The standard technique to reduce the effects of noise in the time-frequency domain is to send the received signal through a bandpass spectral filter and then to implement temporal gating. As depicted schematically in Fig.~\ref{Fig:Filtering}(a), the action of such a sequential incoherent filter (SIF) can be described in terms of the effective number $N$ of temporal modes that make it to the detection stage which is approximately equal to the time-bandwidth product of the filter \cite{Raymer2020}. Correspondingly, the noise strength can be written as:
\begin{equation}\label{eq:nb}
    n_b = N \cdot \frac{\mathscr{N}}{hf_c},
\end{equation}
where $\mathscr{N}$ is the noise power spectral density. In order to ensure that the signal is not significantly attenuated, typically the effective number of modes $N$ is at least of the order of $100$. Novel techniques, such as QPG shown in Fig.~\ref{Fig:Filtering}(b) and discussed in Sec.~\ref{Sec:QPG} may allow to reduce substantially this number without significant signal loss.

\section{PIE limits}
\label{sec:pie_limits}

Let us start by reviewing canonical communication efficiency limits expressed in terms of the PIE. For simplicity we will deal with an optical signal in a single spatial mode that experiences linear attenuation in the course of propagation.

Because the ultimate efficiency limits are determined by the quantum nature of light used as information carrier, particular attention needs to be paid to the mathematical description of signal modulation and demodulation. When the optical carrier is described within the quantum mechanical formalism, signal modulation maps the macroscopic electrical (classical) signal onto the quantum state of light that is subsequently transmitted over the optical (quantum) channel. In the demodulation process, the optical (quantum) signal is converted back into a macroscopic electrical (classical) signal. In the quantum mechanical description, this conversion process is described as a quantum measurement that has its own limitations \cite{DaviesIEEETIT1978,Shapiro2009}.  
In many practical scenarios these limitations are manifested as detection noise. A typical example is coherent detection of one or two optical field quadratures by means of homodyning. Even if the complex amplitude of the incident optical field is defined as well as possible (i.e.\ it is represented in the quantum formalism by a so-called coherent state), the measurement outcome exhibits Gaussian noise whose minimum level for an ideal setup is given by the shot noise limit. In the fully quantum mechanical description of coherent detection, the shot noise limit can be interpreted as a consequence of quantum fluctuations of electromagnetic field quadratures whose level is determined by the Heisenberg uncertainty principle \cite{Yuen1983,ShapiroIEEEJQE1985}.

In general, detection noise depends on a specific measurement scheme and needs to be treated separately from the background noise acquired by the optical signal in the course of propagation through the quantum channel. A canonical model for the latter is the additive white Gaussian noise (AWGN) model, which has a natural extension to the quantum formalism. A convenient figure of merit to characterize the noise strength is the AWGN power spectral density expressed in photon energy units:
\begin{equation}
n_{n} = \frac{\mathscr{N}}{hf_c}.
\label{Eq:nndef}
\end{equation} 
The above quantity describes the average number of background photons acquired by the signal temporal mode. While for a SIF discussed in Sec.~\ref{Sec:PIE} the number of temporal modes reaching the detection stage is typically much larger than one in order to ensure that the power carried by the signal mode is not suppressed, ultimate capacity limits are derived under an assumption that only noise present in the signal mode is detected. Such selectivity is intrinsic to coherent detection. Its physical realization in the case of other detection schemes, e.g.\ photon counting, will be discussed in Sec.~\ref{Sec:QPG}.

When shot noise limited coherent detection is used to detect one (S1) or two conjugate (S2) quadratures of the electromagnetic field, the attainable information rates are given by the Shannon-Hartley theorem\cite{Shannon1949a} and read
\begin{align}
    {\sf R}_{\textrm{S1}} & = \frac{B}{2} \cdot \log_2 \left(1+\frac{4 n_s}{1+2n_n}\right), 
    \label{Eq:RS1def} \\
    {\sf R}_{\textrm{S2}} & = B \cdot \log_2 \left(1+\frac{n_s}{1+n_n}\right), \label{Eq:RS2def}
\end{align}
where $n_s$ and $n_n$ are defined respectively in Eqs.~(\ref{Eq:nsdef}) and (\ref{Eq:nndef}).
The fractions ${4 n_s}/({1+2n_n})$ and ${n_s}/({1+n_n})$ have the straightforward interpretation of the signal-to-noise ratio (SNR).\cite{Banaszek2020} Noteworthily, denominators in these fractions include contributions from the detection noise, equal to 1 in the chosen units, and the background noise, proportional to $n_n$, that enter with different weights, depending on the detection scheme. 

In the limit of unrestricted bandwidth, $B\rightarrow \infty$, the logarithms in Eqs.~(\ref{Eq:RS1def}) and (\ref{Eq:RS2def}) can be expanded up to the linear term in $n_s$. The results recast in terms of the photon information efficiency take the form:
\begin{align}
    {\sf PIE}_{\textrm{S1}} & = \frac{1}{Bn_s} {\sf R}_{\textrm{S1}}  \rightarrow \frac{2}{1+2 n_n } \log_2 e, 
    \label{Eq:PIES1} \\
    {\sf PIE}_{\textrm{S2}} & = \frac{1}{Bn_s} {\sf R}_{\textrm{S2}}  \rightarrow \frac{1}{1+ n_n } \log_2 e \label{Eq:PIES2}
\end{align}
When the background noise is very weak, $n_n \rightarrow 0$, Eqs.~(\ref{Eq:PIES1}) and (\ref{Eq:PIES2}) reduce respectively to ${\sf PIE}_{\textrm{S1}} = 2 \log_2 e \approx 2.88$ and ${\sf PIE}_{\textrm{S2}} = \log_2 e \approx 1.44$ bits per photon. These values stem from the shot noise inherent to conventional quadrature detection and determine limitations of coherent detection to achieve efficient communication in the photon-starved regime.

The ultimate limit on the communication capacity of a quantum channel is given by the Gordon-Holevo (GH) expression \cite{GordonProcIRE1962,Giovannetti2014}
\begin{equation}
\label{Eq:GHCapacity}
    {\sf R}_{\textrm{GH}} = B \cdot [g(n_s + n_n) - g(n_n)],
\end{equation}
where $g(x) = (x+1) \log_2 (x+1) - x \log_2(x)$. This formula incorporates optimization over all modulation formats under a given average power constraint, as well as physically realizable measurements. Importantly, in general it is not possible to identify in the GH expression a single parameter involving the signal strength and the noise strength that could be interpreted as a signal-to-noise ratio. This contrasts with the classical Shannon-Hartley expression. 

\begin{figure}
\begin{center}
\begin{tabular}{c}
\includegraphics[height=5.5cm]{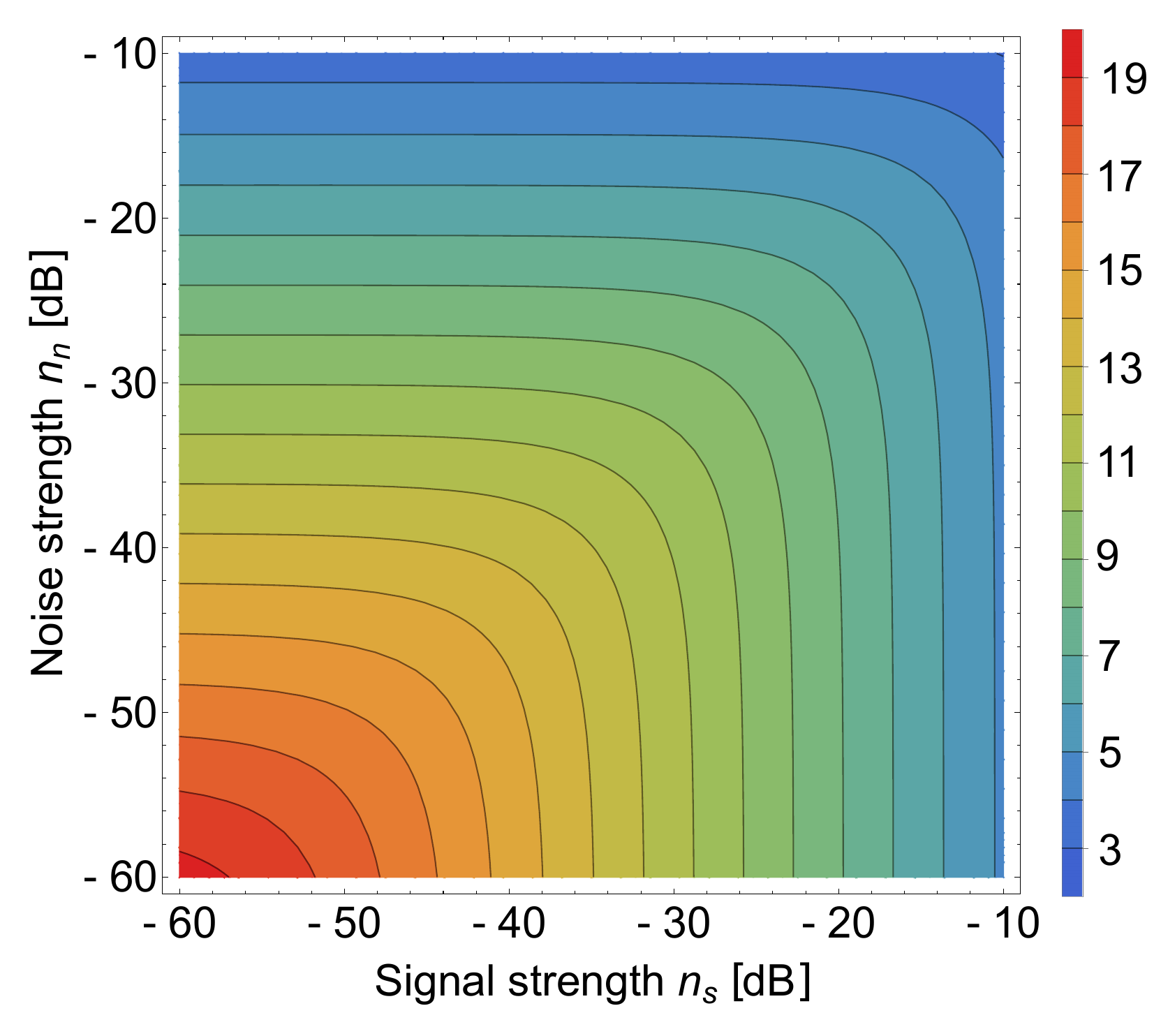}
\end{tabular}
\end{center}
\caption 
{ \label{fig:Holevo}
The Gordon-Holevo limit on the photon information efficiency 
as a function of the signal strength $n_s$ and the background noise strength $n_n$.} 
\end{figure}

The GH capacity bound given in Eq.~(\ref{Eq:GHCapacity}) implies the following limit on the PIE
\begin{equation}
      \textsf{PIE}_{\textrm{GH}}  = \frac{1}{Bn_s} {\sf R}_{\textrm{GH}} = \frac{1}{n_s} [g(n_s + n_n) - g(n_n)] 
\end{equation}
which is plotted as a function of $n_s$ and $n_n$ in Fig.~\ref{fig:Holevo}. 
It is seen that the GH PIE limit increases both with decreasing $n_s$, which corresponds to higher available bandwidth under a fixed signal power, as well as with decreasing background noise $n_b$. 
In particular, in the absence of background noise, $n_n=0$, one obtains
\begin{equation}
  \textsf{PIE}_{\textrm{GH}} = g(n_s)/n_s = \log_2 (1/n_s) + 1 + O(n_s). 
  \label{Eq:PIEGHnb=0}
\end{equation}
This implies that with decreasing $n_s$ (increasing bandwidth) the GH PIE limit can take an arbitrarily high value. However, when the background noise is non-zero, the GH PIE limit is upper bounded by a finite value which is reached in the limit $n_s \rightarrow 0$ \cite{Jarzyna2017, Ding2019}:
\begin{equation}
  \textsf{PIE}_{\textrm{GH}} \rightarrow  g'(n_n) = \log_2 \left(1 + \frac{1}{n_n}\right).
\end{equation}
Inserting this asymptotic formula in the expression for the information rate given in Eq.~(\ref{Eq:RasPIE}) yields 
\begin{equation}
{\sf R}_{\textrm{GH}} \approx \frac{P_{\textrm{rx}}}{hf_c} \cdot \log_2 \left(1 + \frac{hf_c}{\mathscr{N}}\right).
\label{Eq:RGHinfB}
\end{equation}
As expected for power-limited communication, the information rate is proportional to the received signal power $P_{\textrm{rx}}$. Interestingly, the energy $hf_c$ of a single photon at the carrier frequency---and consequently Planck's constant---features explicitly in the above expression. This highlights the relevance of the quantum nature of light to achieve efficient photon-starved communication.

The conventional Shannon limit for power-limited communication is recovered from Eq.~(\ref{Eq:RGHinfB}) when the background noise is strong in the photon energy units, ${\mathscr{N}} \gg hf_c$. In this case, one can expand the logarithm up to the linear term in $hf_c/\mathscr{N}$, which gives the well known classical formula:
\begin{equation}
    {\sf R} \approx  \frac{P_{\textrm{rx}}}{\mathscr{N}} \log_2 e,
    \label{Eq:Rhighnn}
\end{equation}
in which the energy of a single photon no longer explicitly manifests itself. 
The result (\ref{Eq:Rhighnn}) is also obtained from the one- and two-quadrature Shannon limits by neglecting in the denominators of either Eq.~(\ref{Eq:RS1def}) or Eq.~(\ref{Eq:RS2def}) the shot noise contribution, equal to $1$, with respect to the background noise contribution, given respectively by $2n_n$ or $n_n$, and considering the regime $n_s/n_n \ll 1$. This observation implies that conventional coherent detection is optimal when the background noise is strong, which is usually the case of RF communication owing to the presence of the thermal background. As an illustrative example,\cite{Moisson2014} for an RF link with the carrier frequency $f_c=32$~GHz
and the noise power spectral density $\mathscr{N}=-178.45$~dB-mW/Hz one obtains 
${\sf PIE}_{\textrm{RF}} \approx ( hf_c / \mathscr{N} ) \log_2 e \approx 0.0214$ bits per photon, which in Eq.~(\ref{Eq:RasPIE}) gives optical communication an advantage, as in the optical band the PIE can reach values well above one. 

\section{Pulse position modulation}
\label{sec:PPM}

The current standard for photon efficient communication is the pulse position modulation format combined with photon counting direct detection as depicted schematically in Fig.~\ref{Fig:PPM}(a). 
In the idealized scenario with no background noise and no detector dark counts, the only impairment that may occur in signal demodulation is an erasure event, when no click is recorded over the duration of an entire PPM frame. It will be convenient to denote by $n_f$ the mean photon number in the received pulse characterizing its optical energy. Because the pulse occupies one of $M$ otherwise empty slots, one has
\begin{equation}
    n_f = M \cdot n_s. 
\end{equation}
If the pulse photon number statistics is Poissonian, the probability that an erasure has not occurred reads $p= 1 - \exp(-n_f)$ and the amount of information that can be recovered from a single PPM frame is $p \cdot \log_2 M = [1 - \exp(-n_f)]\cdot \log_2 M$. Consequently, the photon information efficiency reads:
\begin{equation}
    {\sf PIE}_{\textrm{PPM}} = \frac{1}{n_f} \cdot [1 - \exp(-n_f)]\cdot \log_2 M.
    \label{Eq:PIEPPM}
\end{equation}
For a fixed PPM order $M$, maximum PIE is achieved in the limit $n_f \rightarrow 0$ and is equal to $\log_2 M$. This is in agreement with an elementary intuition that a single photon positioned in one of $M$ separate temporal slots can carry $\log_2 M$ bits of information.

For a fixed signal strength $n_s$ one can identify the optimal PPM order, or equivalently the photon number per frame $n_f$, which maximizes the PIE. Assuming that $M$ is large enough so that it can be treated as a continuous parameter, the optimal PIE is well approximated by \cite{WangIEEETIT2014,Jarzyna2015}
\begin{equation}
\label{Eq:PIEPPMnonoise}
    {\sf PIE}_{\textrm{PPM}}^\ast = \log_2 (1/n_s ) - \log_2 \log (1/n_s) + O(1). 
\end{equation}
and the optimal pulse optical energy is 
\begin{equation}
    n_f^\ast \approx 2[\log ({2e}/{n_s})]^{-1},
\end{equation}
which implies rather weak dependence of $n_s$ over the range of signal strengths under consideration.
It is instructive to compare the  expression given in Eq.~(\ref{Eq:PIEPPMnonoise}) with Eq.~(\ref{Eq:PIEGHnb=0}). One can notice that in the absence of the background noise, there is a double-logarithmic gap between the efficiency of the optimized PPM/DD combination and the ultimate Gordon-Holevo PIE limit. 

\begin{figure*}
\begin{center}
\begin{tabular}{c}
\includegraphics[width=14cm]{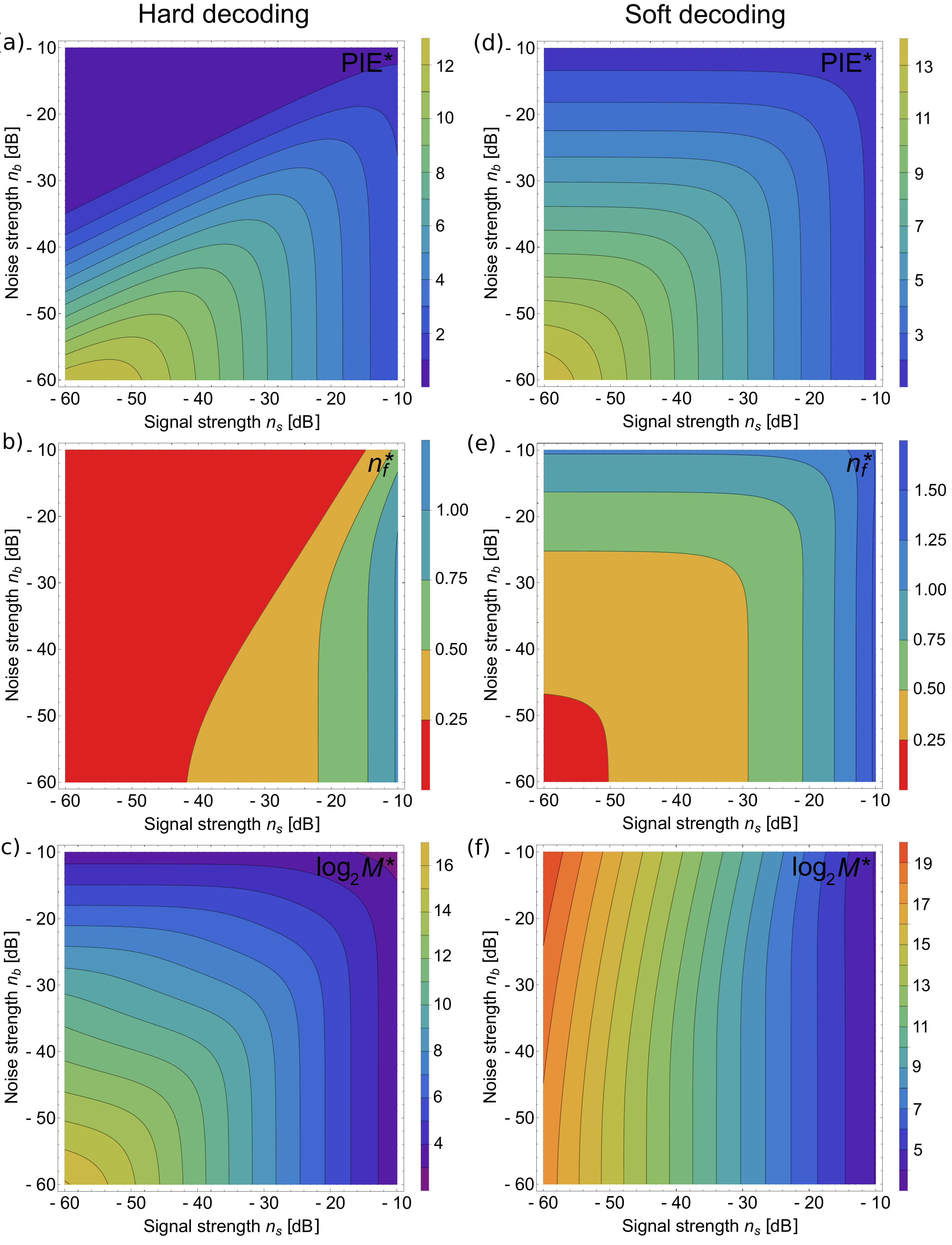}
\end{tabular}
\end{center}
\caption 
{ \label{Fig:G_SIF}
Optimization of the photon information efficiency ${\sf PIE}$ for the PPM format over the pulse optical energy $n_f$ for a given signal strength $n_s$ and noise strength $n_b$ assuming conventional sequential incoherent filtering of the received signal. 
In the hard decoding scenario (a-c) information is retrieved only from PPM frames where only a count in a single slot has been recorded, whereas in the soft decoding scenario (d-f) counts in multiple slots are also processed. 
} 
\end{figure*}

Analysis of the efficiency of the PPM format in the presence of background noise is more intricate, as a detector count in a given temporal slot can be generated either by a pulse, with a probability $p$, or an empty slot, with a non-zero probability $q$, as illustrated in Fig.~\ref{Fig:PPM}(b). When an idealized model of a photon counting detector which does not exhibit dead time, afterpulsing, etc.\ is considered, counts in individual slots are statistically uncorrelated and may occur multiple times in one PPM frame. We shall analyze first the scenario of conventional sequential incoherent filtering shown in Fig.~\ref{Fig:Filtering}(a), which results in multimode background noise at the detection stage.\cite{Raymer2020} Suppose that the filtering subsystem passes to the photon counting detector the total number of $N$ modes, among them the information-carrying signal mode. If the background noise per mode is weak in photon energy units, i.e.\ $\mathscr{N} \ll hf_c$, and the number of modes is large $N \gg 1$, the photocount statistics generated by background noise will be Poissonian with the expectation value $n_b$ given by Eq.~(\ref{eq:nb}). Consequently, the probabilities of generating a photocount respectively by a pulse and an empty slot read
\begin{equation}
    p = 1 - \exp(-n_f - n_b), \quad q = 1 - \exp(-n_b).
    \label{Eq:PPMGeigermultimode} 
\end{equation}
In the simplest approach of hard decoding, information is read out from frames that contain exactly one count, while all other cases are treated as erasures. 
However, such a strategy results in rather poor performance in terms of attainable PIE, 
as seen in Fig.~\ref{Fig:G_SIF}(a-c). Specifically, the attainable PIE  
scales as ${\sf PIE} \sim n_s$ with the vanishing signal strength, $n_s \rightarrow 0$ \cite{Zwolinski2018}.
In the soft decoding approach, information is obtained from all combinations of counts in a PPM frame. Although the calculation of the exact amount of information that can be retrieved under such decoding is rather complicated, there exists a very useful information theoretic bound of the form\cite{HamkinsISIT2004} 
\begin{equation}\label{eq:PIE_lower}
    \textsf{PIE}_{\textrm{PPM}} \ge \frac{1}{n_f} \cdot D\bigl(p|| (n_s/n_f) p + (1-n_s/n_f) q \bigr)
\end{equation}
where $D(x||y)=x\log_2(x/y)+(1-x)\log_2[(1-x)/(1-y)]$ stands for the relative entropy (Kullback-Leibler divergence) between two binary probability distributions $(x,1-x)$ and $(y,1-y)$ and $p$ and $q$ are conditional probabilities shown in Fig.~\ref{Fig:PPM}(b). Conveniently, the bound given in Eq.~(\ref{eq:PIE_lower}) becomes tight in the limit of high PPM orders, i.e.\ when $M = n_s/n_f \ll 1$.
This bound is derived under the assumption that information encoding and decoding can be implemented with information theoretic Shannon efficiency. 

Results of optimization of the right hand side of Eq.~(\ref{eq:PIE_lower}) over the pulse optical energy $n_f$ (or equivalently the PPM order $M$, treated as a continuous parameter) are shown in Fig.~\ref{Fig:G_SIF}(d-f). Special care needs to be taken when comparing the attainable PIE as a function of the noise strength with the GH PIE limit shown in Fig.~\ref{fig:Holevo}. Whereas in the former case the noise figure $n_b$ includes contributions from all the detected modes that make it to the detector through the SIF as defined in Eq.~(\ref{eq:nb}), the GH PIE limit is depicted as a function of noise $n_n = \mathscr{N}/h f_c$ present in the signal mode only. Consequently, if the SIF transmits effectively e.g.\ $N=100$ modes, a given value of $n_b$ in Fig.~\ref{Fig:G_SIF}(a) should be related to $n_n$ in
Fig.~\ref{fig:Holevo} that is $20$~dB lower. 

The results shown in Fig.~\ref{Fig:G_SIF}(d) indicate that for a given noise strength $n_b$ it is beneficial to increase the bandwidth $B$ (thus reducing the signal strength $n_s$) to a level such that 
$n_s \lesssim n_b$. The optimal pulse energy $n_f^\ast$ exhibits weak dependence on either of the parameters $n_s$ or $n_b$ and remains at a level somewhat below one. These observations allow one to identify the efficient operating regime for the PPM/DD combination in the following way. For a given received signal power $P_{\textrm{rx}}$, the duration of one PPM frame should be chosen according to Fig.~\ref{Fig:G_SIF}(e) such that it carries of the order of one photon to generate a click with an appropriately high probability. This frame needs to be divided into a sufficiently high number of slots $M^\ast$ so that the average number of signal photons per slot is below that of the background noise. As seen in Fig.~\ref{Fig:G_SIF}(f) this requires rather high PPM orders. 

\section{Enhanced direct detection receivers}
\label{Sec:Enhanced}
This section will address two possible enhancements of a photon counting receiver for a scalable PPM format. The first one is QPG depicted in Fig.~\ref{Fig:Filtering}(b), which in principle has the capability to filter out in a nearly lossless way the signal mode and to reject the background noise present in other modes that would pass through a conventional SIF. The second potential enhancement is PNR detection, which can improve discrimination of PPM signal pulses from a weak background noise.

\begin{figure*}[htbp]
\begin{center}
\begin{tabular}{c}
\includegraphics[width=14cm]{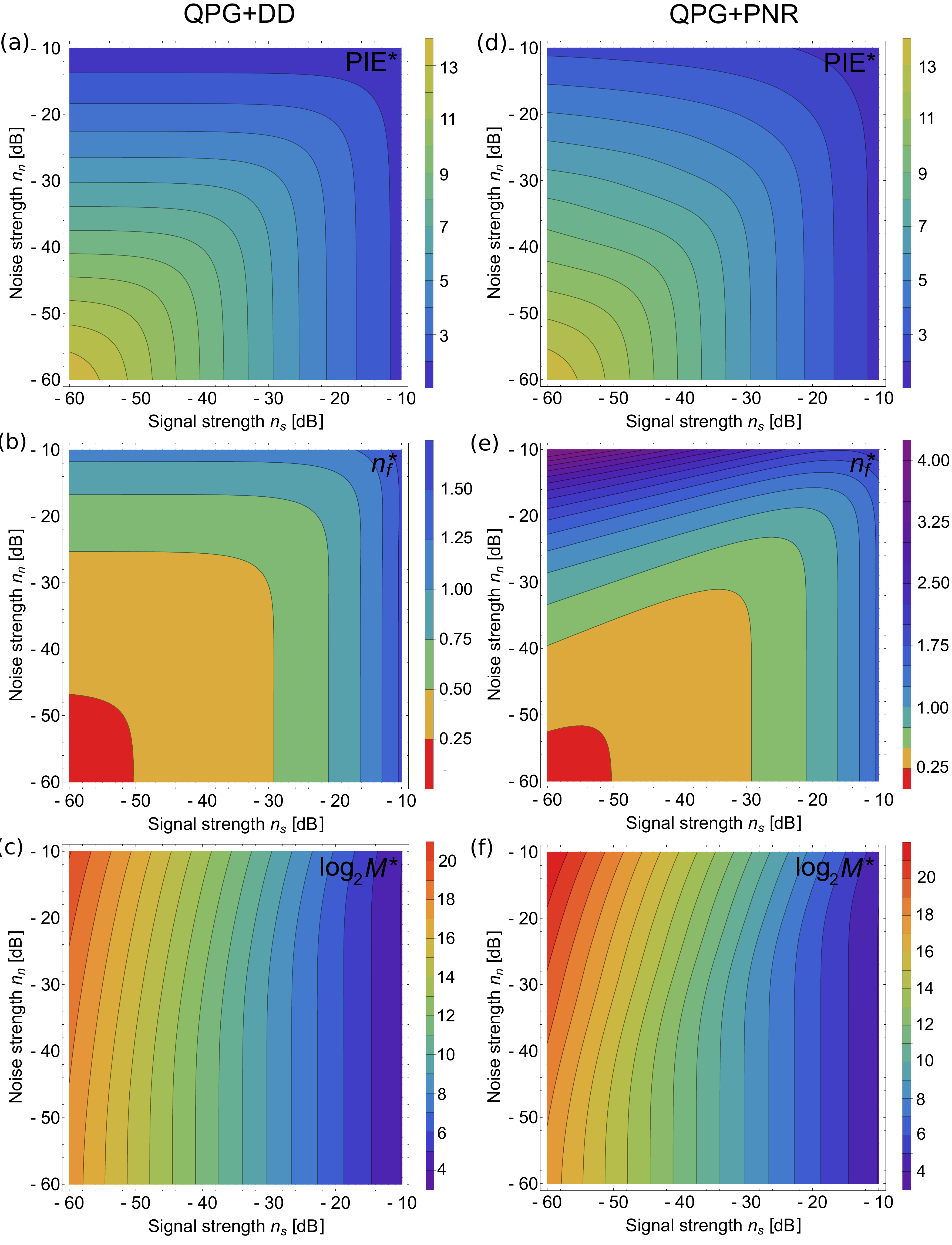}
\end{tabular}
\end{center}
\caption 
{ \label{Fig:EnhancedRx}
Optimization of the photon information efficiency ${\sf PIE}$ for the PPM format over the pulse optical energy $n_f$ for a given signal strength $n_s$ and noise strength $n_n$ with a quantum pulse gating (QPG) filter for the received signal. Conventional binary on-off detection (a-c) is compared with photon number resolved (PNR) detection (d-f).} 
\end{figure*} 

\subsection{Quantum pulse gating (QPG)}
\label{Sec:QPG}

The basic idea of QPG is shown schematically in Fig.~\ref{Fig:Filtering}(b). 
The signal beam, containing multiple temporal modes, is combined in a $\chi^{(2)}$ non-linear optical medium with a pump pulse to realize the sum frequency generation process. With an appropriate choice of the pump pulse shape and the medium phase matching function, only the signal temporal mode is upconverted to the sum-frequency band, whereas all other temporal modes containing background noise remain at the input carrier frequency. In principle, this technique allows one to reject noise present in modes orthogonal to the signal one, thus implementing an all-optical matched filter for detection techniques other than homodyning or heterodyning.

When calculating the conditional probabilities depicted in Fig.~\ref{Fig:PPM}(b) one needs to take into account that the amplitude of the PPM pulse exhibits Gaussian fluctuations owing to the acquired background noise. This leads to the following formulas for generating a count by a pulse and an empty slot \cite{Marian1993}:
\begin{equation}
p = 1-\frac{1}{1+n_n}\exp\left(-\frac{n_f}{1+n_n}\right) , \quad q = \frac{n_n}{1+n_n}.
\label{Eq:PPMGeigersinglemode}
\end{equation}
Optimization of the right hand side of Eq.~(\ref{eq:PIE_lower}) with the above expressions over the pulse optical energy $n_f$ yields results shown in Fig.~\ref{Fig:EnhancedRx}(a-c). Because the noise strength refers now only to noise present in the signal mode, the noise values $n_n$ parametrizing the vertical axes relate directly to those used in Fig.~\ref{fig:Holevo}. Qualitatively, the behavior of the optimized PIE, the pulse optical energy $n_f^\ast$, and the PPM order $M^\ast$ is analogous to the scenario considered in Sec.~\ref{sec:PPM}. However, it should be kept in mind that the QPG noise rejection mechanism will reduce $n_n$ parameterizing vertical axes in Fig.~\ref{Fig:EnhancedRx} with respect to $n_b$ used in Fig.~\ref{Fig:G_SIF} by $(10\log_{10} N)$~dB, where $N$ is the effective number of modes transmitted through a SIF.

\subsection{Photon number resolved (PNR) detection}

The detection model considered so far assumed that each slot produces a binary outcome in the form of a count when at least one photon has been registered, or a no count at all. More generally, the detector can have photon number resolving capability and return the actual integer number $k=0,1,2,\ldots$ of registered photons. If a PNR detector is preceded by a QPG filter, the statistics of the count number $k$ for a pulse and an empty slot are respectively as follows \cite{Marian1993}:
\begin{align}
   p_k & = \frac{1}{1+n_n} \exp\left( -\frac{n_f}{1+n_n} \right) 
   L_k \left(-\frac{n_f}{n_n(1+n_n)}\right), \nonumber \\
   q_k & = \frac{1}{1+n_n}\left( \frac{n_n}{1+n_n} \right)^k,
\end{align}
where $L_k(\cdot)$ denotes the $k$th Laguerre polynomial. In order to calculate attainable PIE, the relative entropy on the right hand side of Eq.~(\ref{eq:PIE_lower}) needs to be taken 
for distributions of an integer variable $k=0,1,2,\ldots$ rather than a binary variable indicating whether a count has occurred in a given slot or not. Optimization of this expression over the pulse optical energy $n_f$ yields results shown in Fig.~\ref{Fig:EnhancedRx}(d-e). Interestingly, one can notice improvement in the attainable PIE when $n_s \ll n_n$. This is associated with a qualitatively different behavior of the optimal pulse optical energy $n_f^\ast$ in this regime. Whereas in Fig.~\ref{Fig:EnhancedRx}(b) $n_f^\ast$ tended to a finite value with $n_s \rightarrow$ for a fixed background noise level $n_n$, it is seen that in the PNR case $n_f^\ast$ does not seem to be upper bounded.

The above observations have the following physical interpretation in the regime of weak background noise, when $n_n \ll 1$. If the detector has photon number resolving capability it is beneficial to increase the optical energy of the PPM pulse so that it generates with a substantial probability events corresponding to the detection of two or more photons. These can be discriminated against events generated by background noise, which owing to its low level, will correspond predominantly to detection of single photons. An advanced mathematical analysis of the PIE limit shows that for a given noise strength $n_n$, the GH PIE limit is indeed saturated for unrestricted bandwidth, when $n_s \rightarrow 0$ \cite{Jarzyna2021, Jarzyna2023}. However, one should note that convergence is exceedingly slow and requires even higher PPM orders than those indicated previously, as seen in Fig.~\ref{Fig:EnhancedRx}(f).

\section{Discussion}
\label{Sec:Discussion}

A common feature for photon-starved communication strategies considered in this paper is the qualitatively different behavior of the PIE depending on the relation between the signal strength $n_s$ defined in Eq.~(\ref{Eq:nsdef}) and the background noise strength, given by either Eq.~(\ref{eq:nb}) or Eq.~(\ref{Eq:nndef}) as determined by the implemented noise rejection mechanism.  As long as the background noise is much weaker that the signal, the PIE can be improved by increasing the signal bandwidth (slot rate) $B$, albeit the scaling is logarithmic: as seen from Eq.~(\ref{Eq:PIEGHnb=0}) and Eq.~(\ref{Eq:PIEPPMnonoise}), a two-fold increase in the signal bandwidth (halving the slot duration) adds only one bit of information to the attainable PIE. When the background noise strength exceeds $n_s$, the PIE value saturates at a finite level that is a function of the noise strength. 
This asymptotic value ${\sf PIE}^{\ast\ast}$ can obtained from Eq.~(\ref{eq:PIE_lower})
by taking the limit of vanishing signal strength, $n_s \rightarrow 0$, and performing one-parameter optimization \cite{Verdu1990, Zwolinski2018}
\begin{equation}
{\sf PIE}^{\ast\ast} = \sup_{n_f \ge 0} \frac{1}{n_f} D(p||q).
\end{equation}
The above expression determines the photon information efficiency attainable without any restrictions on the signal bandwidth.

\begin{table}[t]
\caption{Comparison of the expected downlink data rates from the Psyche mission to the ground receiver at the Aristarchos Telescope\cite{RielanderICSO2022} with attainable information rates for unrestricted bandwidth in scenarios considered in this paper: SIF+DD, sequential incoherent filter and soft-decoded binary on/off direct detection; QPG+DD, quantum pulse gating filter and binary on/off direct detection; Gordon-Holevo, the ultimate quantum mechanical limit. The nighttime background noise strength assumed in ``Night'' columns is $n_b = -39.5$~dB for the SIF+DD scenario and $n_n = -72.5$~dB in the remaining two cases. The latter figure is increased to $n_n = -42.5$~dB for the hypothetical daytime operation scenario in ``Day'' columns.} 
\label{Tab:DataRates}
\begin{center}       
\begin{tabular}{|c|c|c|c|c|c|c|c|c|} 
\hline
Distance & Signal flux & Expected  & \multicolumn{6}{c|}{Attainable information rate [Mbps]} \\
{[AU]}    & [photons/s] & data rate  & \multicolumn{2}{c|}{SIF + DD} & \multicolumn{2}{c|}{QPG + DD} 
& \multicolumn{2}{c|}{Gordon-Holevo} \\
         &             & [Mbps] & Night & Day & Night & Day & Night & Day \\
\hline
$1.25$ & $1.67\times 10^5$ & $0.456$  & $1.435$   &  ---  & $3.087$  &  $1.579$  & $4.023$   & $2.359$   \\
\hline
$1.50$ & $1.16\times 10^5$ & $0.456$  & $0.997$   &  ---  & $2.144$  &  $1.097$  & $2.794$   & $1.638$   \\
\hline
$1.75$ & $8.53\times 10^4$ & $0.228$  & $0.733$   &  ---  &  $1.577$ &  $0.807$  & $2.055$   & $1.205$   \\
\hline
$2.00$ & $6.53\times 10^4$ & $0.228$  & $0.561$   &  ---  & $1.207$  &  $0.617$  & $1.573$   & $0.922$   \\
\hline
$2.25$ & $5.16\times 10^4$ & $0.228$  & $0.443$   &  ---  & $0.954$  &  $0.488$  & $1.243$   & $0.729$   \\
\hline
$2.50$ & $4.18\times 10^4$ & $0.114$  & $0.359$   &  ---  & $0.773$  &  $0.395$  & $1.007$   & $0.590$   \\
\hline
$2.75$ & $3.45\times 10^4$ & $0.114$  & $0.296$   &  ---  & $0.638$  & $0.326$   & $0.831$   & $0.487$   \\
\hline
\end{tabular}
\end{center}
\end{table} 

Theoretical limits obtained in the preceding sections can serve as a benchmark for the performance of concrete deep-space optical communication systems. An interesting example is the Psyche mission equipped with a laser communication terminal that is planned to transmit data to a ground receiver based on the Aristarchos Telescope in Greece. Following downlink budget calculations presented by Riel\"{a}nder {\em et al.},\cite{RielanderICSO2022} the receiver filter bandwidth $2$~nm and the slot duration $8$~ns imply that the effective number of modes detected in a single slot by a conventional DD receiver is approximately $N \approx 2000$. A typical night time background photon flux $1.40 \times 10^{4}$ photons per second yields for an $8$~ns slot the background noise strength $n_b = -39.5$~dB, which after QPG filtering would reduce to $n_n = -72.5$~dB. The attainable information rates derived from PIE limits for three scenarios considered in the preceding sections are given in columns labeled ``Night'' in Table~\ref{Tab:DataRates} and compared with the data rates expected for the actual system. The column ``Day'' refers to a hypothetical daytime operation of the link based on an assumed $30$~dB penalty in the noise strength.\cite{Hemmati2005} It is seen that the use of the QPG noise rejection technique could enable operation of deep-space optical communication links also in daytime conditions at data rates that are comparable with or higher than those attainable using SIF at night time.

\begin{figure}
\begin{center}
\begin{tabular}{c}
\includegraphics[width=15cm]{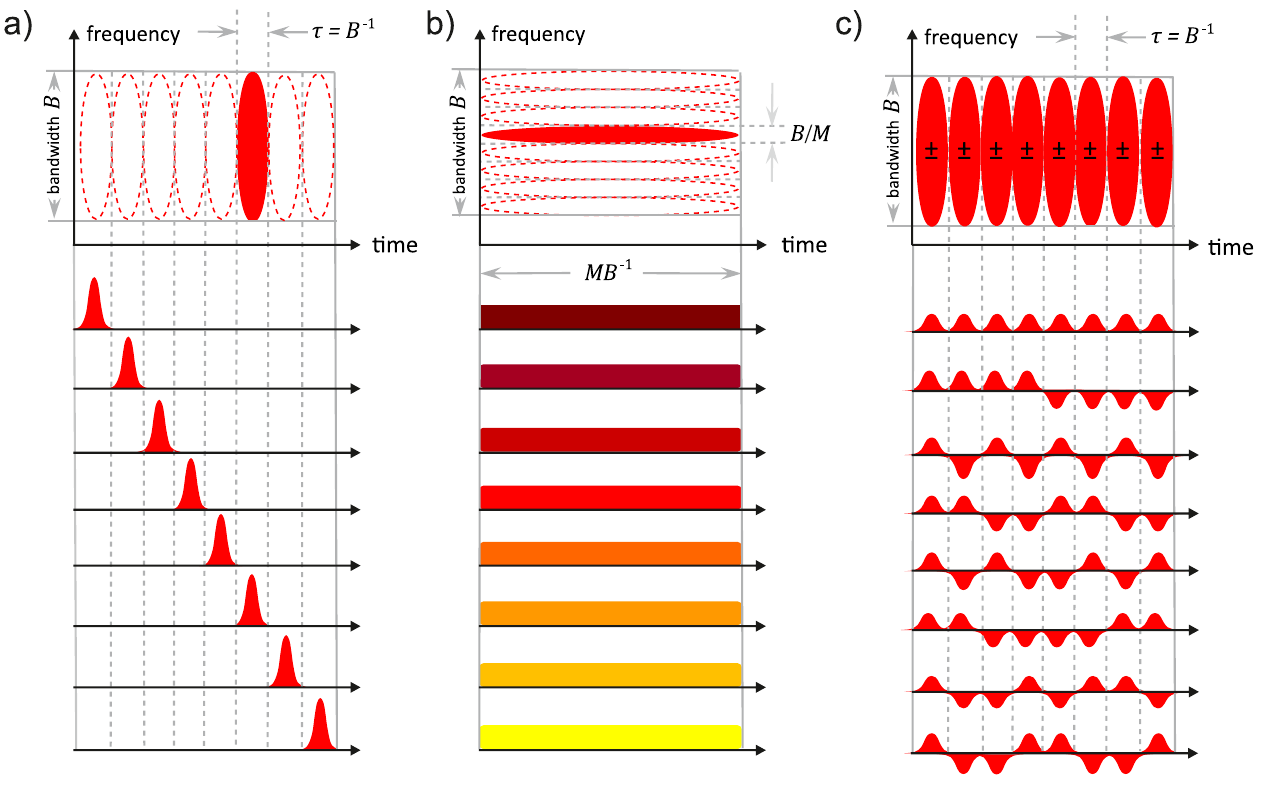}
\end{tabular}
\end{center}
\caption{
Scalable modulation formats visualized using time-frequency diagrams. For a format order $M$, a single symbol occupies an area determined by a bandwidth $B$ on the frequency axis and a duration $MB^{-1}$ on the time axis. This area can be sliced in the temporal domain, which corresponds to the PPM format (a), or in the spectral domain which produces the frequency shift keying (FSK) format shown in (b). More generally, one can define a set of $M$ mutually orthogonal temporal modes that occupy the  $MB^{-1} \times B$ time-frequency area as illustrated in the panel (c) with Hadamard words composed from the BPSK alphabet.
\label{Fig:ScalableModulationFormats}} 
\end{figure} 

Reaching high PIE values requires a scalable modulation format that can be implemented at high orders. In the case of PPM, communication with frames of length up to $2^{15}$ slots has been demonstrated in laboratory settings,\cite{FarrChoi2013} and very recently the PPM order $2^{19}$ has been used to achieve the PIE of 12.5 bits per photon with independent free-running clocks at the transmitter and the receiver.\cite{EssiambreXXX2023} Generation of a high-order PPM signal for photon-efficient communication poses a number of technical challenges, such as managing high peak-to-average power ratio, and ensuring a sufficiently good extinction ratio so that empty slots do not carry any residual optical radiation. As illustrated in Fig.~\ref{Fig:ScalableModulationFormats}, analogous PIE values can be achieved with other modulation formats,\cite{BanaszekJachuraICSO2018} such as frequency shift keying (FSK),\cite{SavageOpEx2013} or words composed from the binary phase shift keyed (BPSK) alphabet\cite{GuhaPRL2011} and demodulated with multistage interferometric receivers,\cite{BanaszekJachuraICSOS2017,ZwolinskiECOC2020,CuiXXX2023} where technical challenges would take different forms. 

\section{Conclusions}
\label{Sec:Conclusions}

This paper discussed theoretical limits on the performance of deep-space optical communication links in terms of photon information efficiency. The presented results can serve as a benchmark for actual realizations. Without bandwidth restrictions, the key limitation is the background noise acquired by the propagating signal. The effects of background noise can be reduced by novel noise rejection techniques, such as quantum phase gating based on carefully engineered nonlinear optical interactions. 
Approaching the photon information efficiency limit for a given background noise strength requires a modulation format that can be scaled up to very high orders. While so far the pulse position modulation format has been successfully employed in both practical systems and laboratory demonstrations, at some point it may encounter technical barriers, such as limits on the achievable peak-to-average power ratio that will make it necessary to revisit other options for scalable modulation formats. 

\section*{Acknowledgments}

We wish to acknowledge insightful discussions with R.-J. Essiambre, S. Halte, C. Heese, C. Marquardt, M. G. Raymer, C. Silberhorn, and M. D. Shaw. This work was a part of the project ``Quantum Optical
Technologies'' carried out under the International Research Agendas Programme
of the Foundation for Polish Science co-financed by the European Union through
the European Regional Development Fund. W. Z.'s contribution was supported from Polish budget funds for science in the years 2019–2023 as a research project under the “Diamond Grant” programme, contract no. 0118/DIA/2019/48.

\section*{Code, Data, and Materials Availability}

All data in support of the findings of this paper are available within the article.

\bibliography{report}   
\bibliographystyle{spiejour}   

\end{document}